\documentclass[preprints,article,accept,moreauthors,pdftex]{Definitions/mdpi} 

\firstpage{1} 
\pubvolume{1}
\issuenum{1}
\articlenumber{0}
\pubyear{2023}
\copyrightyear{2023}
\externaleditor{Academic Editor: Nuno Silva}
\datereceived{30 August 2023} 
\daterevised{20 September 2023} 
\dateaccepted{21 September 2023} 
\datepublished{date} 
\hreflink{https://doi.org/} 

\Title{Open-Source 
	 Magnetometer for Characterizing Magnetic Fields in Ultracold Experiments}

\TitleCitation{Open-Source Magnetometer for Characterizing Magnetic Fields in Ultracold Experiments}

\Author{Koray 
 Din{\c {c}}er and Mariusz Semczuk *\orcidA{}}

\AuthorNames{Koray Din{\c {c}}er and Mariusz Semczuk}

\AuthorCitation{Din{\c {c}}er, K.; Semczuk, M.}

\address[1]{ Institute of Experimental Physics, University of Warsaw, Pasteura 5, 02-093 Warsaw, Poland;} 
\corres{\hangafter=1 \hangindent=1.05em \hspace{-0.82em}Correspondence:  msemczuk@fuw.edu.pl}

\abstract{
We demonstrate a fully automated open-source magnetometer designed primarily for characterization of magnetic fields produced by coils, permanent magnets or by parasitic sources. It is based on an Arduino Mega microcontroller and a three-axis Hall sensor with a measurement range of $\pm$8~G per axis and the RMS of the field readout below 0.3~mG. For all practical purposes, the sensor displacement during data acquisition is virtually unlimited, which can be particularly useful for characterizing large or extended coils like Helmholtz cages or Zeeman slowers.
All components needed for the construction are cheap and widely available off-the-shelf elements or are 3D-printed. The operation of the magnetometer is controlled via a graphical user interface (GUI), which manages all essential functionalities, like data acquisition and plotting. The GUI also incorporates additional features, like data averaging, calibration of the displacement of the Hall sensor or real-time readout of the magnetic field, useful for monitoring magnetic field changes.
We have used a pair of rectangular coils constructed for a potassium--cesium 2D magneto-optical trap to benchmark the performance of the magnetometer. We have obtained good agreement with both simulations and measurements acquired with a commercial gaussmeter.}


\begin{document}

\section{\label{sec:introduction}Introduction}
Magnetic fields are ubiquitous in ultracold experiments. They are generated either by the flow of current through an arrangement of conductors (called coils)~\cite{PhysRevLett.48.596,PhysRevLett.54.2596,PhysRevLett.77.416} or by permanent magnets~\cite{PhysRevA.51.R22,OVCHINNIKOV2007261}. Their role can vary from enabling optical trapping in magneto-optical traps~\cite{PhysRevLett.59.2631}, control of interactions with Feshbach resonances~\cite{RevModPhys.82.1225} and assisting with slowing of atomic velocities in Zeeman slowers~\cite{10.1063/1.4945567} to setting a quantization axis and nullifying stray magnetic fields~\cite{app112110474}.

When constructing a new experimental setup or upgrading an existing one with an additional source of magnetic field, one relies on simulations to guide the final design. This usually provides a fairly good estimation of what magnetic fields and their gradients one should expect within the region of interest, such as over the extent of a laser-cooled cloud of atoms.  In~practice, simulations rarely deviate from measurements by more than several percent~\cite{thesis:msc_pawel}, but~this can be said only in hindsight
, after~the magnetic field source has been characterized. Typical coils used in ultracold experiments are home-built and the winding process uses equipment constructed ad~hoc to guide and secure a relatively thick (often 1--5~mm in diameter) copper wire to assure well-organized layers. In~the end, however, windings rarely form a nicely organized stack as assumed in simulations. Mounting coils in the setup might also introduce tilts and cause deviations from the simulated arrangement. Permanent magnets, on~the other hand, are harder to simulate reliably without any prior knowledge of the field generated by each individual magnet, and it is essential to actually measure the magnetic field produced by the final~assembly.

In ultracold experiments, magnetic fields need to be changed, often quickly ($\sim$1~\textmu s), during~various stages of the experimental sequence. This flexibility can be easily achieved with coils, which makes them ubiquitous in many experimental setups. Even in a very simple single-chamber setup used to create a magneto-optical trap, at least four pairs of coils are present: one pair to create a quadrupole potential plus three perpendicularly oriented pairs to provide a uniform magnetic field for stray fields compensation and setting the quantization axis. This number can easily grow for dual-species experiments with an independent source for each atom type~\cite{PhysRevA.84.011601}, containing multi-stage Zeeman slowers~\cite{10.1063/1.4945567} or using an Ioffe--Pritchard-type magnetic trap~\cite{PhysRevLett.96.020401}. The~prevalence of coils in modern setups would suggest that there is abundance of commercial or home-made tools to characterize magnetic fields. Commercial gaussmeters are widely available, but they cannot be used right out of the box as they are usually handheld (like, e.g., Hirst GM07 used in this work), which requires the user to build some sort of a rig that would enable sliding the sensor in and out of the region one wants to characterize while noting the position of each magnetic field readout. In~practice, such measurements can be time-consuming, and, to partially compensate for that, the spatial resolution or the measurement range are often decreased. Somewhat surprisingly, there are very few reports on home-built magnetometers that would provide functionality and ease of implementation matching our design. What can be found is mentioned in passing in various master's or doctoral theses~\cite{thesis:msc_pawel, ospelkaus2006fermi}, and even then these solutions seem to offer a rather limited measurement range, with~virtually no discussion of the performance. As~such, the~designs are virtually impossible to reproduce by others, and this forces many researchers to construct their own tools. Some of the solutions found in the literature use microcontrollers like Arduino to read out data from a Hall sensor, but they often offer very little in terms of functionality. They lack fully automatic measurements~\cite{GTorzo_1987,Atkin_2016,Ishafit_2020,10.1119/5.0044498}, rely on closed solutions like National Instruments data acquisition cards and Labview~\cite{Ishafit_2020,10.1119/5.0044498} or do not possess a general sensor-moving mechanism that is compatible with characterization of large coils, as commonly found in ultracold physics experiments~\cite{BALOGUN2018443}.  

In this work, we present a detailed design of an open-source magnetometer that is based on cheap and
widely available off-the-shelf elements and 3D-printed components, facilitating easy reproduction. All the necessary designs and code are provided and can be found in a GitHub repository~\cite{automag}. The~operation of the device is controlled via a graphical user interface that contains all the necessary functionalities that enable fast and accurate characterization of magnetic fields with minimal human supervision. The~magnetometer can be constructed and operated by a person with only very basic experience with electronics and with virtually no experience in~programming.

We discuss how one can measure not only the magnetic fields produced by the source (coils, permanent magnets) but also the stray fields present in the laboratory. We present features implemented in the software, like automated calibration, which determines the displacement of the sensor per revolution of the stepper motor or the real-time readout of the magnetic field. The~performance is illustrated with a measurement of a magnetic field produced by a pair of coils designed for a potassium and cesium 2D magneto-optical~trap.

\section{\label{sec:general_consideration}General~Considerations}

Whenever a field $\vec{B}$ is measured with a magnetometer, one needs to remember that, in a typical laboratory environment, it is in~fact a~vector sum of the field generated by the source and of stray fields. The~latter ones can be shielded to a large degree with a $\mu$-metal enclosure or otherwise compensated with auxiliary coils, but this is rarely completed during characterization of magnetic field sources prior to assembly. As~a result, straightforward methods to remove the stray fields during the characterization process increase the quality of the measurement.
Let us first consider how one could eliminate stray fields when characterizing a pair of coils. Here, one effectively deals with sets of current-carrying loops. Let us limit our discussion to axially symmetric coils commonly used in magneto-optical traps or in Zeeman slowers, but~the conclusions are~general.

From the Biot--Savart law, we know that the magnetic field vector at a point a distance $y$ along the center line of the circular loop with radius $R$ can be expressed as 

\begin{equation}
    \vec{B}(y\hat{\mathbf{y}})=\frac{\mu_0IR^2}{2(y^2+R^2)^{3/2}}\hat{\mathbf{y}},
    \label{eq:B-S_law}
\end{equation}
where $I$ is the current, $\mu_0$ is magnetic permeability of free space and $\hat{\mathbf{y}}$ is the unit vector along the center-line of the loop centered at the origin. From~this equation, one can see that the direction of the vector changes when the direction of the current is~reversed. 

A magnetometer measures magnetic field at a given point in space, expressed as $\vec{B}_+=\vec{B}_{\mathrm{source}}+\vec{B}_{\mathrm{stray}}+\vec{B}_{\mathrm{offset}}$, where $\vec{B}_+$ corresponds to the current flowing in (arbitrarily chosen) positive direction, $\vec{B}_{\mathrm{source}}$ is the magnetic field generated by the coils, $\vec{B}_{\mathrm{stray}}$ is the stray magnetic field and $\vec{B}_{\mathrm{offset}}$ is an internal offset field inherent to most Hall sensors. If~we reverse the current in the coils, then the magnetometer will measure a new field value $\vec{B}_-=-\vec{B}_{\mathrm{source}}+\vec{B}_{\mathrm{stray}}+\vec{B}_{\mathrm{offset}}$. These measurements can then be used to obtain the field created by the source:
\begin{equation}
    \vec{B}_{\mathrm{source}}=\frac{\vec{B}_+-\vec{B}_-}{2}.
\end{equation}

Alternatively, 
 one could measure the field with the source present and with the source removed (or the current turned off). In~both scenarios, it is impossible to extract the value of the stray fields without the knowledge of the inherent offset of the magnetic sensor. If~the current direction can be changed, then each component of stray fields, for~example $\vec{B}^x_{\mathrm{stray}}$, could be determined by physically rotating the sensor by 180$^{\circ}$ around a perpendicular axis such that new values $\vec{B}^{r,x}_+=-\vec{B}^x_{\mathrm{source}}-\vec{B}^x_{\mathrm{stray}}+\vec{B}^x_{\mathrm{offset}}$ and $\vec{B}^{r,x}_-=\vec{B}^x_{\mathrm{source}}-\vec{B}^x_{\mathrm{stray}}+\vec{B}^x_{\mathrm{offset}}$ are additionally measured. From~the above relations, one can obtain $\vec{B}^x_{\mathrm{stray}}$ from equations 
\begin{subequations}
\begin{eqnarray}
\vec{B}^x_{\mathrm{stray}}=\frac{\vec{B}^x_- -\vec{B}^{r,x}_+}{2}
\\
\vec{B}^x_{\mathrm{stray}}=\frac{\vec{B}^{r,x}_- - \vec{B}^x_+}{2}.
\end{eqnarray}
\end{subequations}

This is very cumbersome and, if one is interested in all three components of the magnetic field vector, the complexity of a measurement would become significant. As~such, it is rarely conducted because stray field measurements are usually treated as supplementary information of no relevance to the testing process of the coils. In~fact, it is rather common to even ignore the presence of the stray fields, and whatever the magnetometer measures is treated as the field created solely by the source. For~permanent magnets, the~best one can do is to measure the magnetic field when the magnets are present and then repeat the measurement after the magnets are physically removed such that they do not influence the stray field value. In~general, however, the~above discussion does not provide a solution to a question that in some instances might be important: how to reliably characterize the field from a source that cannot be removed or turned off, as~is the case with stray fields. 

\subsection*{Field Measurement with Internal Offset~Cancellation\label{subsec:int_offs_cancel}}
Nowadays, many magnetic field sensors are available for do-it-yourself projects, in~particular as peripherals for Arduino and the like. Very commonly encountered technology is based on the Hall effect. The~magnetic fields detected by such sensors are prone to temperature drifts because of an internal offset field inherent to the device. Some sensor models like MEMSIC MMC5983MA~\cite{memsic-datasheet} used in our design have a mechanism that effectively flips the magnetic sensing polarity of the sensing element. A~simplified procedure determines the sensor's response to the external magnetic field by performing measurements for two polarizations of the sensor, $\vec{B}_{\mathrm{pol+}}=\vec{B}+\vec{B}_{\mathrm{offset}}$ and $\vec{B}_{\mathrm{pol-}}=-\vec{B}+\vec{B}_{\mathrm{offset}}$~\cite{memsic-datasheet}. As~a result, from~the above equations, one can obtain both the external field and the offset. The~external field $\vec{B}$ could be just stray magnetic field already present at the measurement location, usually Earth's magnetic~field. 

To measure the field from a source, $\vec{B}_{\mathrm{source}}$, one needs to remember that the sensor's readout $\vec{B}_\mathrm{sensor}$ is composed of three fields (as already defined in this section), $\vec{B}_\mathrm{sensor}=\vec{B}_{\mathrm{source}}+\vec{B}_{\mathrm{stray}}+\vec{B}_{\mathrm{offset}}$. We can utilize polarization flipping feature to extract all needed fields with the following~procedure:
\begin{enumerate}
    \item without the magnetic field source, measure the sensor's response at each location $y_i$ along the planned measurement path for two polarizations of the sensor, $\vec{B}^i_{\mathrm{ns,+}}=\vec{B}^i_{\mathrm{stray}}+\vec{B}^i_{\mathrm{offset}}$ and $\vec{B}^i_{\mathrm{ns,-}}=-\vec{B}^i_{\mathrm{stray}}+\vec{B}^i_{\mathrm{offset}}$.
    \item place the magnetic field source and repeat the measurement for the same locations $y_i$ as before, for~two polarizations of the sensor, $\vec{B}^i_{\mathrm{ws,+}}=\vec{B}^i_{\mathrm{source}}+\vec{B}^{'i}_{\mathrm{stray}}+\vec{B}^{'i}_{\mathrm{offset}}$ and $\vec{B}^i_{\mathrm{ws,-}}=-(\vec{B}^i_{\mathrm{source}}+\vec{B}^{'i}_{\mathrm{stray}})+\vec{B}^{'i}_{\mathrm{offset}}$,
\end{enumerate}
where $\vec{B}^i_{ns/ws,+/-}$ are sensor readouts with no source (ns) and with the source present (ws) for two polarities (+/$-$). In~principle, the~values marked with the prime symbol are not the same as the ones without the prime symbol because they are obtained many seconds apart. We assume here, however, that the stray fields are static on the time scale of the measurement and we do not distinguish between the two. We do not make this assumption about the offset field as this can be influenced by the fields the sensor has been exposed to. We have observed such variation in the sensor offset (see Section~\ref{sec:measure_fields}); therefore, the measurement of both $\vec{B}^i_{\mathrm{ws,+}}$ and $\vec{B}^i_{\mathrm{ws,-}}$ is necessary to reliably extract the field characterizing the source, the~environment and the sensor: 
\begin{subequations}
\label{eq:fields_perm_magn}
\begin{eqnarray}
\vec{B}^i_{\mathrm{source}}&=&\frac{\vec{B}^i_{\mathrm{ws,+}}-\vec{B}^i_{\mathrm{ws,-}}-\vec{B}^i_{\mathrm{ns,+}}+\vec{B}^i_{\mathrm{ns,-}}}{2},
\label{eq:fields_perm_magn_1}
\\
\vec{B}^i_{\mathrm{stray}}&=&\frac{\vec{B}^i_{ns,+}-\vec{B}^i_{ns,-}}{2},
\label{eq:fields_perm_magn_2}
\\
\vec{B}^{'i}_{\mathrm{offset}}&=&\frac{\vec{B}^i_{\mathrm{ws,+}}+\vec{B}^i_{\mathrm{ws,-}}}{2},\\
    \label{eq:fields_perm_magn_3}
\vec{B}^i_{\mathrm{offset}}&=&\frac{\vec{B}^i_{\mathrm{ns,+}}+\vec{B}^i_{\mathrm{ns,-}}}{2}.
    \label{eq:fields_perm_magn_4}
\end{eqnarray}
\end{subequations}

When dealing with coils, we can improve upon the procedure discussed above. This comes from the possibility of turning off the current and thus obtaining the stray field value within tens of milliseconds from the readout of the magnetic field when the coils are on. The~overall procedure is very similar to the case of permanent magnets, and we use the same naming convention for the fields. The~acquisition procedure is maximally sped up if the measurements are performed in the following order:

\begin{enumerate}
    \item while the coil current is off, measure the sensor's response at location $y_i$ along the planned measurement path for two polarizations of the sensor, $\vec{B}^i_{\mathrm{ns,+}}=\vec{B}^i_{\mathrm{stray}}+\vec{B}^i_{\mathrm{offset}}$ and $\vec{B}^i_{\mathrm{ns,-}}=-\vec{B}^i_{\mathrm{stray}}+B^i_{\mathrm{offset}}$,
    \item turn on the coil current and repeat the measurement for the same locations $y_i$ as before for~two polarizations of the sensor, $\vec{B}^i_{\mathrm{ws,+}}=\vec{B}^i_{\mathrm{source}}+\vec{B}^{'i}_{\mathrm{stray}}+\vec{B}^{'i}_{offset}$ and $\vec{B}^i_{\mathrm{ws,-}}=-(\vec{B}^i_{\mathrm{source}}+\vec{B}^{'i}_{\mathrm{stray}})+\vec{B}^{'i}_{\mathrm{offset}}$.
    \item while keeping the coil current on, repeat the measurement for a new location $y_{i+1}$, for~two polarizations of the sensor, $\vec{B}^{(i+1)}_{\mathrm{ws,+}}=\vec{B}^{(i+1)}_{\mathrm{source}}+\vec{B}^{(i+1)}_{\mathrm{stray}}+\vec{B}^{(i+1)}_{\mathrm{offset}}$ and $\vec{B}^{(i+1)}_{\mathrm{ws,-}}=-(\vec{B}^{(i+1)}_{\mathrm{source}}+\vec{B}^{(i+1)}_{\mathrm{stray}})+\vec{B}^{(i+1)}_{\mathrm{offset}}$.
    \item turn off the coil current and repeat as in step 1 but now for the new location $x_{i+1}$.
    \item move the sensor to the location $y_{i+2}$ and cycle through steps 1 to 4 until the final point in the planned measurement path is reached.
\end{enumerate}

The un-primed and primed values are measured much closer in time if compared to the permanent magnet case; therefore, much faster changes in stray fields (e.g., due to running elevators creating additional magnetic field) are acceptable. The~field of interest can be calculated using Equation~(\ref{eq:fields_perm_magn}).


\section{Experimental Setup}
\unskip

\subsection{\label{subsec:hardware}Hardware}

The choice of hardware solutions has been driven by the desire to make the entire magnetometer cheap and easy to assemble while providing high performance in terms of achievable noise and resolution of the magnetic field measurement. We have taken care to make our design easy to reproduce. The~mechanical assembly uses easily available commercial components like T-slotted aluminum profiles or threaded steel rods and 3D-printed parts that can be made in-house. All the 3D-printer-ready designs are available in a GitHub repository~\cite{automag}.

\begin{figure}[H]
\includegraphics[width=\columnwidth]{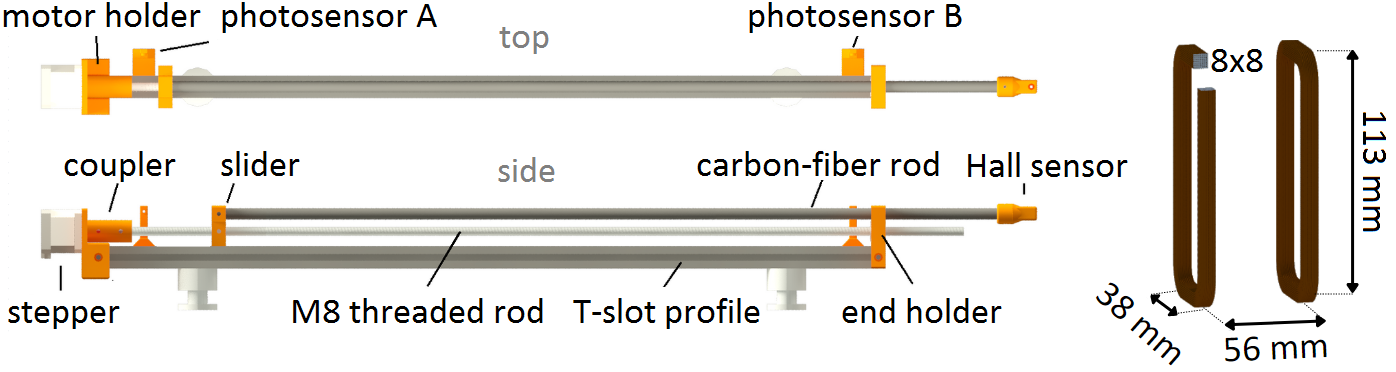}
\caption{\label{fig:mechanical_schematic} The layout of the automated magnetometer as seen from the top and from the side. Orange-colored parts are 3D-printed. The~rotation of the stepper motor shaft is coupled to the threaded rod via a coupler. An M8 nut is embedded in the slider to convert the rotation of the rod into the translational motion of the carbon fiber tube, which supports the Hall sensor. The~end holder uses a ball bearing to assure smooth rotation of the rod. The~photosensors establish the minimum and maximum range of the motion to avoid collisions with the supporting structure in case the information about the current position of the sensor is lost. The~dimensions of the pair of coils (1~mm wire, 8 $\times$ 8 turns) used for testing of the device are~shown.}
\end{figure}

The friction between 3D-printed elements and the aluminum profile is negligible, and we achieve a smooth one-dimensional motion of the Hall sensor. In~our approach, the~measurements are completed after the motion of the sensor has stopped, so any vibrations or shaking that could take place during the motion itself do not noticeably influence the readout. The~aluminum profile is the ''backbone'' of the setup and everything is attached to it (Figure~\ref{fig:mechanical_schematic}), but neither the profile nor the details of the remaining components need to follow our recommendations;  as long as the general approach we present is followed, the remaining hardware and software will be fully~usable.

The measurement range is set by the length of the T-slot profile and lengths of accordingly chosen carbon tube and the threaded rod. The~range is limited by the acceptable degree of deflection of the extended carbon tube, but, if some form of a supporting structure placed close to the Hall sensor is added, the~effects of deflection could be minimized. Our implementation currently offers a 50~cm range because this is sufficient to measure the field between a pair of coils in a Helmholtz configuration that are used in our laboratory for stray magnetic field compensation. However, for~some long Zeeman slowers~\cite{var_pitch_zeeman_slower,10.1063/1.3600897,PhysRevA.90.063607}, the ability to measure magnetic field over a distance exceeding 1~m is highly~beneficial.

The magnetometer uses a Hall sensor attached via an adapter to a non-magnetic tube made of carbon fiber, commonly used in building model airplanes. We guide wires connecting the sensor to the microcontroller inside of it. An M8 nut embedded in the slider block enables conversion of the rotation of the threaded steel rod into the translational motion of the tube (Figure~\ref{fig:mechanical_schematic}). The~threaded rod itself is connected to the shaft of the stepper motor with a coupler. This allows precise manipulation of the sensor's position in one dimension. Smooth rotation of the threaded rod is assured by a ball bearing embedded in the end holder. We have not implemented any active position feedback that would track the location of the Hall sensor, and we rely on the microcontroller to count the steps of the stepper motor. We have implemented photosensors that define the minimum and the maximum allowed position of the sensor such that the stepper motor would always move within the pre-defined range. We can imagine a realistic scenario when the user fine-tunes the location of the Hall sensor by rotating the threaded rod by hand, which, effectively, would make the microcontroller lose the information about its current location. The~photosensors serve as a safety switch guaranteeing that there will be no collisions with the end holder or with the stepper~motor.

The longitudinal resolution per step of the automated magnetometer $\Delta{y}={P_{\mathrm{M8}}/{N_{\mathrm{step}}}}$ depends on the pitch of the threaded rod, $P_{\mathrm{M8}}$, and~on the number of steps per revolution of the stepper motor, $N_{\mathrm{step}}$. The~stepper motor we use (\mbox{JK42HM48-1684}, driven by a dual H-Bridge motor driver, L298
) has 400 steps per revolution and the rod has a standard M8 metric thread with a pitch of 1.25~mm, enabling the theoretical longitudinal resolution $\Delta{x}=3.125\,$$\upmu$m/step. 

We recommend determining $\Delta{y}$ using the \textit{Calibrate} feature of our software such that measuring the pitch of the thread can be avoided. Here, the~slider moves to one of the photosensors, stops and asks the user to mark the initial position of the slider. Next, the~slider moves towards the other photosensor while the number of steps is counted by the control software. Once the slider has stopped, the~user is asked to measure the displacement from the initial position and to enter that displacement in a pop-up window. The~software calculates $\Delta{y}$ (in our case, 3.123~$\upmu$m/step) and saves the value in a configuration file such that it can be used for future measurements. The~entire device uses the metric system, but~we emphasize that it is, effectively, unit blind; i.e.,~if the separation is entered in inches, everything will be measured in imperial units, even though the GUI would show metric units. The~calibration procedure does not rely on anything particular regarding the pitch of the threaded rod. We use a standard metric M8 thread for convenience, even though the resolution could be more than doubled with a metric M8x0.5 thread. Since we envision the main application of the magnetometer to be for characterizing macroscopic coils, increasing resolution even further provides no added benefit.

 The Hall sensor is the key component of our magnetometer as~it is responsible for detecting magnetic fields. To~ensure accurate measurements, we have used a commercially available Hall sensor chip (MMC5983MA) integrated with an electronic board. The~MMC5983MA is a 3-axis, 18-bit Hall sensor board with an operational range of $\pm$8~G and a total RMS noise of 0.4~mG. It has an internal degaussing function, which is used to remove any residual magnetization that may arise, such as from~overexposure (when the magnetic field exceeds $\pm$8~G), and is realized by the Hall sensor by sending a 500~ns current pulse upon receiving a SET command. Once overexposed, every single measurement that follows cannot be trusted; therefore, degaussing is performed before every readout such that, once the field drops within the sensor's range, the~measured values become reliable again and one does not lose data. The~RESET function has a similar role to the SET function, but~it also flips the polarization of the sensor. Each single measurement is composed of four commands: SET (degaussing), MEASURE (read out bit values from the sensor), RESET (flip the polarization of the sensor and degauss) and MEASURE (read out bit values from the sensor that is now oppositely polarized). This approach removes thermally induced offset errors and eliminates residual magnetization caused by strong external fields (Equation (\ref{eq:fields_perm_magn})), particularly useful if the sensor has been saturated during the~measurement.

To asses how good the noise performance of the sensor actually is, we have measured the magnetic field 1000 times after shielding the sensor with $\mu$-metal and we have obtained the RMS noise equal to 0.19~mG, 0.23~mG and 0.27~mG for the $x$, $y$ and $z$-axis, respectively. Alternative Hall sensors with degaussing and polarization flipping features that are compatible with the $\mathrm{I^2C}$ protocol can be implemented in our automated magnetometer (e.g., \mbox{MMC5633NJL-B}, which extends the measurement range to $\pm 30$~G). 

We distinguish two measurement modes that we call the permanent mode (for permanent magnets characterization) and the coil mode (for characterization of coils). In~the permanent mode, we assume certain extension of permanent magnets and we first measure the stray field in that region without magnets present. We then repeat the measurement with the permanent magnets in place. In~the coil mode, the coils do not need to be removed because we use a MOSFET-based power controller switch (ST1168 model, a microcontroller-compatible power switch based on N-channel MOSFET IRF520N~\cite{irf520n-datasheet} by Infineon) 
 to turn on and off the current flowing through the coil. This feature is implemented in our software, so the switching is automated. Here, the~power switch turns off the current flowing through the coils for each measurement point to obtain the stray magnetic field (Equation~(\ref{eq:fields_perm_magn_2})), and then the switch turns on and the magnetic field generated by the coils is measured (Equation~(\ref{eq:fields_perm_magn_1})). The~permanent mode can be used to characterize coils as well if the user turns on the current flow once the stray fields have been measured. Under~normal conditions, we do not expect significant differences between the two approaches, but they are distinctive if one looks carefully: in the coil mode, the determination of the stray field and the determination of the field from the coils in each location are temporally separated by about 200~ms, whereas, for the permanent mode, this could be even hundreds of seconds. This may play a role if stray fields are time-dependent, e.g.,~due to activities of other people in the~laboratory. 
 
\begin{figure}[H]
\centering
\includegraphics[width=0.7\columnwidth]{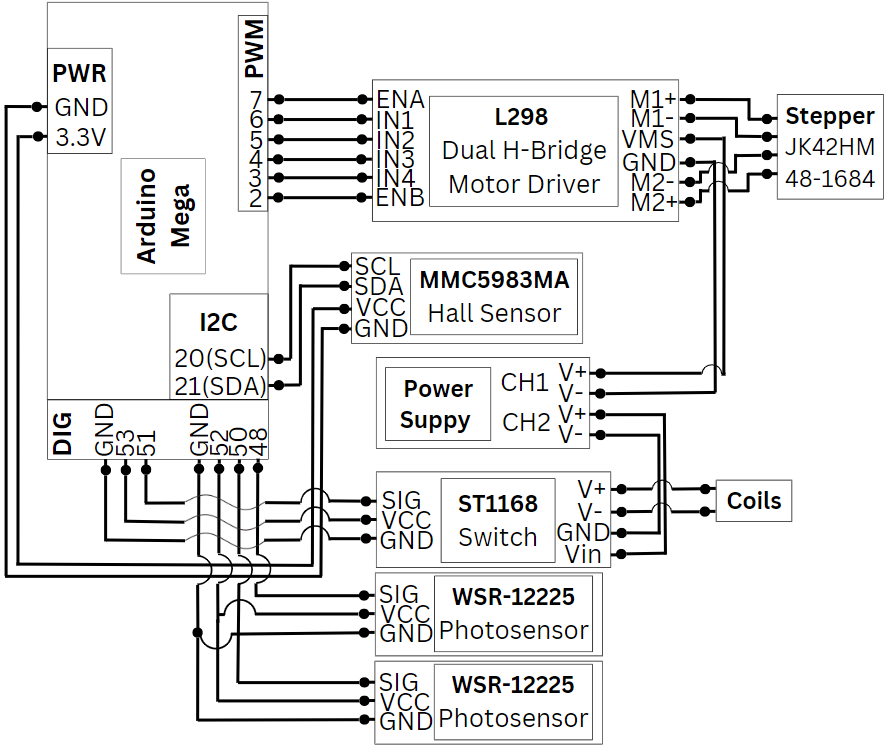}
\caption{\label{fig:electronic_schematic} Schematic of the connections for the electronic components of the automated magnetometer. The~Arduino Mega is controlled by a computer via a USB serial~connection.}
\end{figure}

The electronic components of the automated magnetometer are managed by an Arduino Mega single-board microcontroller (see Figure~\ref{fig:electronic_schematic} for the schematics). Typically, microcontrollers are programmed with firmware designed to carry out specific operations, which can pose a challenge if modifications to the setup are desired, particularly in the case of open-source projects. To~overcome this limitation, we have chosen to use a generic firmware based on the Firmata protocol, which facilitates communication between the software running on a computer and the Arduino board. With~the Firmata firmware uploaded onto the microcontroller, we use APIs (application programming interfaces) to issue commands for executing the desired tasks. We utilize 'Telemetrix-AIO'~\cite{Telemetrix-AIO}, a~firmware based on the generic Firmata protocol, which not only supports all Firmata features but~also includes a stepper motor functionality and communication protocols like $\mathrm{I^2C}$, commonly used by~sensors. 


\subsection{\label{subsec:software}Software}

The 'Telemetrix' firmware has its API package built for Python~3 programming language. 
By~using the API functions in this package, one can send commands to the Arduino for executing specific tasks. The~firmware supports synchronous and asynchronous programming. Asynchronous operations are considered to be an advanced programming method for performing tasks concurrently without blocking the execution of other running tasks. By~utilizing techniques such as event loops and callbacks, asynchronous programming enables the program to initiate tasks and continue execution without waiting for their completion. This concurrent execution approach improves resource utilization and responsiveness, especially in~situations involving waiting for external events or performing I/O operations. In~synchronous programming, tasks are executed in a sequential manner, where each task must finish before the next one can begin. This can introduce delays if a task takes a significant amount of time to complete as~subsequent tasks have to wait their turn. Sensors that only work with $\mathrm{I^2C}$ protocol can be difficult to integrate into a program that requires fast readouts. The~$\mathrm{I^2C}$ protocol can be described as a query-based protocol. In~$\mathrm{I^2C}$ communication, the~master device initiates communication by sending a query or command to the slave device. The~slave device then responds with the requested data or performs the requested action. In~synchronous programming, the~requested data are often handled with the threading method, where a thread of the processor is assigned to report request data. Asynchronous programming, characterized by the efficient utilization of event loops, offers significant advantages in scalability compared to synchronous threading. Event loops, such as the one used in asynchronous programming, facilitate the execution of multiple tasks efficiently by minimizing resource usage and reducing task creation time. In~our specific case, these benefits can be leveraged in the implementation of functions related to the Hall sensor and stepper motor driving. This is particularly important when dealing with long-running operations, such as I/O operations or network~requests. 

To compare the readout time for synchronous and asynchronous programming, we measured the time required to retrieve readouts. Therefore, after~the first readout, a~second one is obtained with a flipped sensor polarization to calculate the offset introduced by the sensor itself (see Equation~(\ref{eq:fields_perm_magn_3})). We have collected 10,000 data points (20,000 readouts in total) for each mode to compare the two programming approaches and have determined that the average time per single data point is 186~ms for synchronous programming and 93~ms for asynchronous programming. These results indicate that the use of asynchronous programming significantly reduces the time required for data retrieval in real-time applications that benefit from fast readouts while guaranteeing a high level of flexibility in implementation of alternative hardware, like a different Hall sensor~model.

\subsection{\label{subsec:gui}Graphical User~Interface}

We have developed a graphical user interface (GUI) that can be used to control all implemented functionalities of the magnetometer~\cite{automag}, effectively removing the need for any programming experience on the user side. The~GUI takes care of plotting the components of both the stray magnetic field and the magnetic field created by magnets. It also enables users to monitor the magnetic field in real time with a rate of approximately 5~Hz. This functionality can be useful, such as when aligning the sensor in the desired orientation or ensuring that it passes through the center of both coils within its range of movement. Simultaneously, it serves as a convenient tool for assessing the stray magnetic field and its fluctuations at the location of the setup. We have utilized asynchronous programming for the GUI as well. With~this approach, GUI updates can occur when a non-blocking task is called, eliminating the need to wait for the command execution to complete. This, in~turn, removes the requirement for multiprocessing to handle GUI operations, which would otherwise involve secure variable exchange between multiple threads using queue operations. By~adopting this approach, we have effectively mitigated potential challenges that may arise when users attempt to edit or implement new features in the~project.

\section{\label{sec:measure_fields}Measurement of Magnetic~Fields}

In the following discussion, as a source of a magnetic field, we use a pair of square coils (see Figure~\ref{fig:mechanical_schematic} for dimensions) wound from a 1~mm wire. This enables us to present the performance for the Helmholtz and anti-Helmholtz arrangement of the magnetic field. The~coils have been designed and assembled to produce an elongated quadrupole field for a 2D magneto-optical trap of cesium and potassium. The~coils are powered by a low-noise two-channel power supply Hameg HMP2020. The~same power supply is also used for powering the stepper motor via its secondary output~channel.

\begin{figure}[H]

\includegraphics[width=.95\columnwidth]{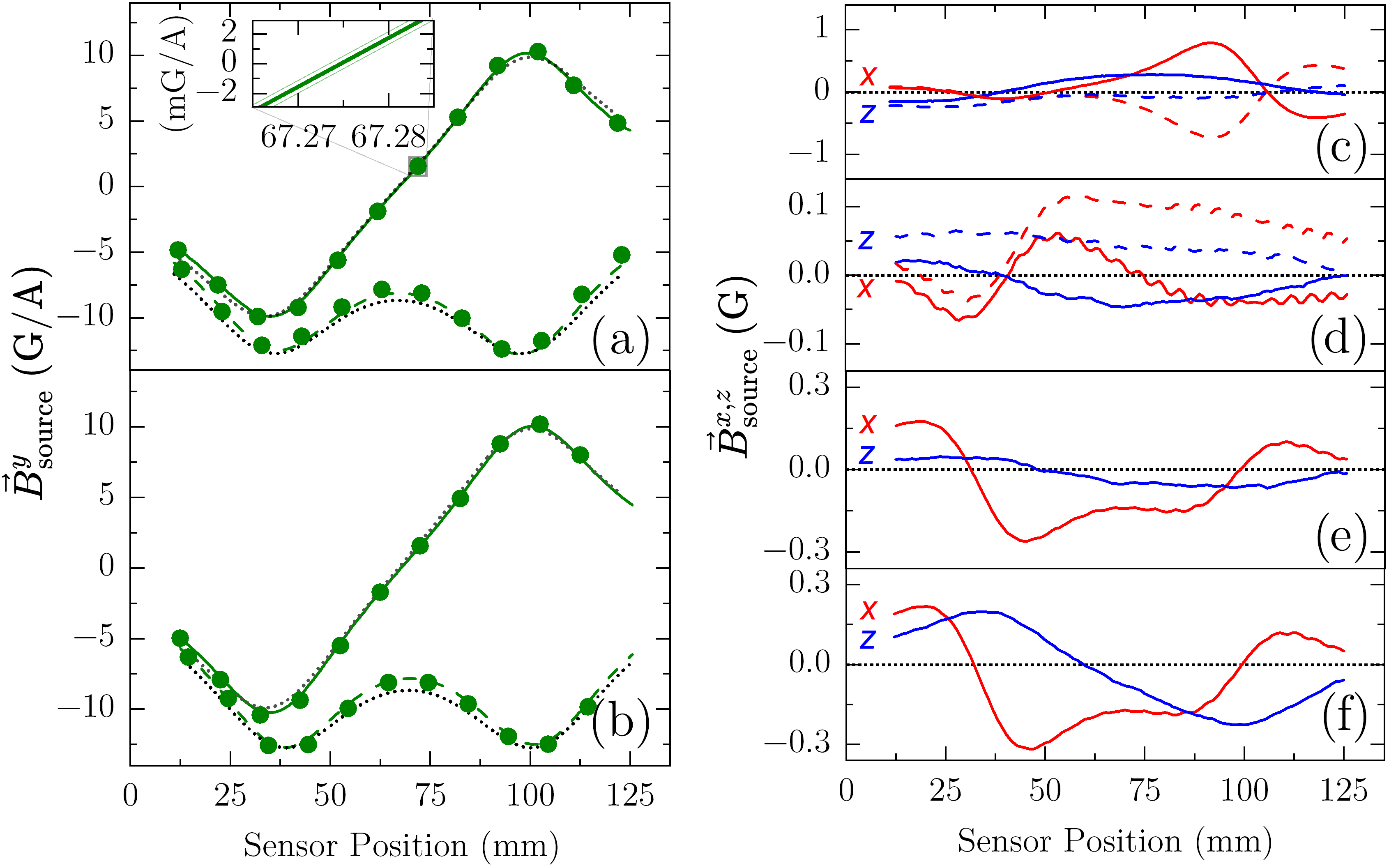}
\caption{\label{fig:coil_xyz} Magnetic 
 field mapping of a set of coils shown in Figure~\ref{fig:mechanical_schematic} operated in a Helmholtz configuration (dashed lines) and anti-Helmholtz configuration (continuous line) measured with the permanent mode with 30-point averaging. The~data in (\textbf{a},\textbf{c}) correspond to the coils assembled in the anti-Helmholtz configuration to operate as a source of a quadrupole field for a 2D magneto-optical trap; (\textbf{b},\textbf{d}) coils individually aligned to minimize the deviations from the ideal anti-Helmholtz configuration; (\textbf{e}) relative arrangement of coils as in (\textbf{d}), with~the assembly rotated by 1~degree around the $z$ axis; (\textbf{f}) relative arrangement of coils as in (\textbf{d}), with~the assembly rotated by 1$^{\circ}$ around the $z$ axis and 1.5$^{\circ}$ around the $x$ axis. The~inset in (\textbf{a}) shows the confidence intervals set by the root mean square value of the measured data points. Magnetic field projections along the sensor axes $x$, $y$ and $z$ are color-coded with red, green and blue, respectively. The~gravity points along the $z$ axis and the $y$ direction are along the translation displacement of the Hall probe. The~green dots in (\textbf{a},\textbf{b}) are measurements obtained with a commercial gaussmeter, as discussed in the main text. We compare our measurements with a simulation of the magnetic field created by the coils, shown with a black dotted line. Here, the~simulation does not take into account the misalignment of the coils with respect to each other nor with respect to the measurement~path.}
\end{figure}

First, we set a reference for the automated magnetometer data by performing a magnetic field measurement with a commercial gaussmeter, Hirst GM07 (Figure~\ref{fig:coil_xyz}a,b). The~observed accuracy of this device is on the order of 1\% of the measured value, in~line with the specifications of the manufacturer. We attach the gaussmeter's sensor to the carbon fiber rod of the constructed magnetometer and use the stepper motor to control its position along the $y$ direction. The~null field adjustment functionality has been used to compensate the stray field at each position. The~magnetic field has been read out from the device's display and written down for each location of the sensor. The~entire procedure to cover a 20~cm distance with a 5~mm spatial resolution can take up to 30~min to~complete. 

To demonstrate the capabilities of the constructed magnetometer, we acquire data using the permanent mode. The~values of interest are determined from the measurements using Equation~(\ref{eq:fields_perm_magn_1}) for the magnetic field generated by the source, Equation~(\ref{eq:fields_perm_magn_2}) for stray magnetic field and Equations~(\ref{eq:fields_perm_magn_3}) and~(\ref{eq:fields_perm_magn_4}) for the offset value of our Hall sensor. These expressions apply to both the permanent mode and the coil mode of the~magnetometer.

\begin{figure}[H] 
\centering
\includegraphics[width=.98\columnwidth]{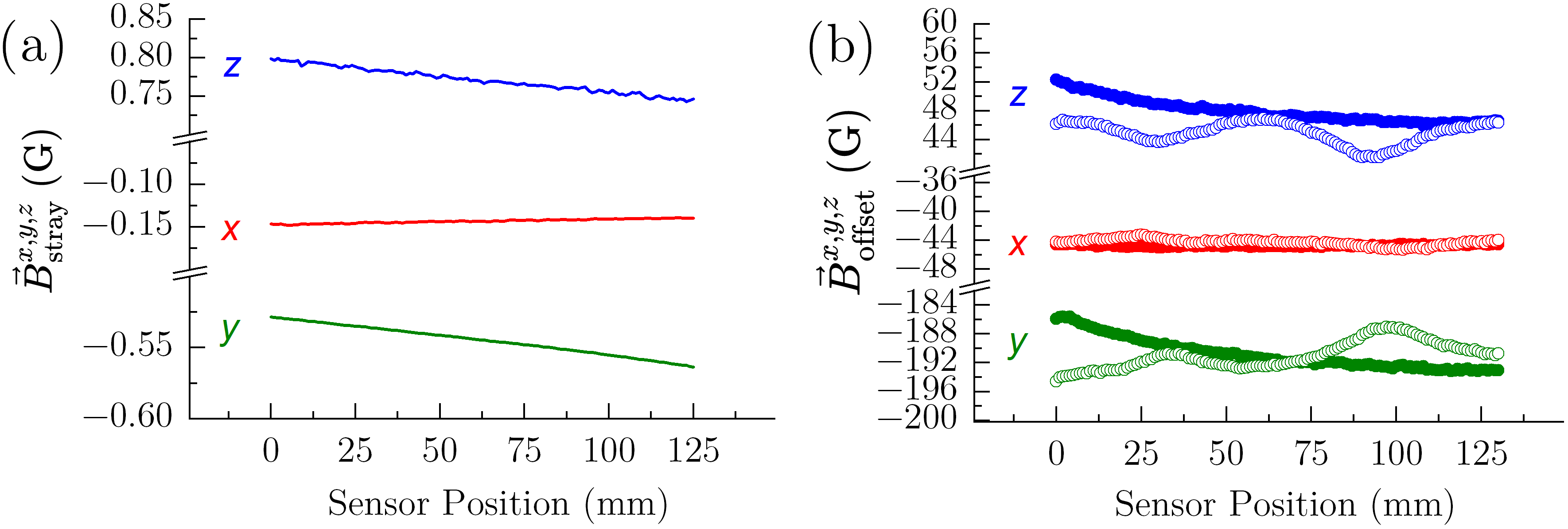}
\caption{\label{fig:stray_offset} (\textbf{a}) Stray 
 field and (\textbf{b}) Hall sensor offset extracted from the same dataset as the one used to calculate the magnetic field of an anti-Helmholtz configuration shown in Figure~\ref{fig:coil_xyz}a,c. In~(\textbf{b}), full circles (empty circles) correspond to the offset determined when the coils are turned off (turned on). Each data point is an average of 30 measurements. Magnetic field projections along the sensor axes $x$, $y$ and $z$ are color-coded with red, green and blue, respectively.}
\end{figure}

In the permanent mode, we start with the MOSFET-based power switch turned off to block the current flowing through our coils (if a magnet was characterized, we would remove it from the setup). We drive the stepper motor by a user-specified range with equally spaced steps until the end point is reached. After~each step, magnetic field readouts are obtained, one for each direction of the sensor polarization. This procedure can be repeated $N$ times for each location to allow averaging. When the sensor reaches its final position, the user is reminded to turn on the power supply for the coils or place the magnet into the setup. With~the completed sequence, we can obtain the stray field values (Figure~\ref{fig:stray_offset}a) and the sensor offset (Figure~\ref{fig:stray_offset}b) at each position of the sensor for all three axes ($x$, $y$ and $z$). Once the user has confirmed the requested operation, the~MOSFET-based power switch is turned on to let the current flow through the set of coils (this is redundant for the permanent magnets). The~stepper motor is then driven in the reverse direction and again for each position (always corresponding to a position where stray fields have been measured), and magnetic field readouts are obtained, one for each direction of the sensor polarization, until~the Hall sensor reaches the initial (starting) position of the sequence. It is worth noting that, in the permanent mode, we observe a backlash of about 0.3~mm due to the change in the direction of motion. This issue is not present when the measurements are performed in the coil mode as the change in the direction of motion does not take place. In~this second run, the~Hall sensor readouts are equal to the sum of the magnetic field generated by the set of coils, the~stray magnetic field and the sensor offset. Equation~(\ref{eq:fields_perm_magn_1}) and the data obtained with the coils off let us extract the $x$,$y$ and $z$ components of the magnetic field, as can be seen in Figure~\ref{fig:coil_xyz}a,c. The~latter figure shows a non-negligible variation in the $x$ and $z$ components of the magnetic field. We have minimized these variations before recording the data used to prepare Figures~\ref{fig:coil_xyz}a,c and~\ref{fig:stray_offset} by adjusting the orientation of the coil assembly to identify the axis of symmetry of the coils. In~the ideal case, the~field along the $x$ and $z$ axes would be zero. We believe that the main reason for the observed asymmetries is that the coils are not exactly the same: they became slightly deformed during winding and are not ideally parallel. All these factors are hard to control when winding coils in house. In~reality, however, these small differences are rarely an issue in actual~experiments.

To characterize sets of assembled coils, one would, ideally, perform measurements by moving the sensor also along the remaining axes. For~this purpose, the~ability to monitor all three components of the magnetic field is unnecessary if the translation is possible only in one direction. However, thanks to the modularity of the presented design, it can be rather easily extended by adding two additional stepper motors to enable full three-dimensional magnetic field mapping and by modifying the part of the code that is responsible for the stepper motor control. This would not require any further modifications to the remaining parts of the software nor the~hardware. 

The ability to measure the magnetic field along all three axes can be, however, used for assessing how well a set of coils has been assembled, and this information could guide adjustments if deemed necessary. We illustrate this by reassembling the coils to minimize the variation in the field along axes $x$ and $z$, which are perpendicular to the direction of the sensor displacement. We run the current through the first coil already attached to the coils mount and adjust its orientation to maximize the $y$ component of the field while minimizing the values along the remaining axes. The~coil mount is then secured in place, and the~sensor is moved (via GUI) to the expected location of the center of the second, which is then aligned in the exact same way as the first one. In~this step, the~current flows exclusively through the second coil. Here, the~range of adjustment is set by the dimensions of the coil holder. This procedure is straightforward, with the real-time mode accessible via GUI. The~data shown in Figure~\ref{fig:coil_xyz}b,d have been obtained after the aforementioned alignment procedure and show significantly smaller variation along $x$ and $z$ than for the original assembly. The~apparent wobbling of the $x$ component of the field is caused by wear of the 3D printed slider, most likely caused by a slightly bent threaded rod. This issue has been fixed by tightening the screw that attaches the slider to the aluminum profile. The~wobbling has been, however, repeatable such that the standard deviation of data for each measurement location differs from the average by less than 5~mG. 

We have investigated the influence of moving the sensor a little off the center of the coils by rotating the assembly first by $1^{\circ}$ around the $z$ axis, which caused an increase in the magnetic field variation along the $x$ axis (Figure~\ref{fig:coil_xyz}e), and then by additionally rotating it by $1.5^{\circ}$ around the $x$ axis, also introducing increased variation along the $z$ axis (Figure~\ref{fig:coil_xyz}e). Under~these rotations, the~magnetic field gradient along the $y$ direction, a~figure of merit for most ultracold experiments, changed by no more than 1\%. It is worth noting that the alignment of the coils that minimized the $x$ and $y$ components caused the decrease in the gradient produced by the coils by about 3\% with respect to the original assembly characterized in Figure~\ref{fig:coil_xyz}a,c.      

In the coil mode, similar to the permanent mode, the~measurements are performed at equally spaced locations for two polarizations of the sensor. At~each point, the~MOSFET-based power switch is initially turned off to block the current flowing through the coils. The~stray magnetic field and the sensor's offset for that location are extracted with Equations~(\ref{eq:fields_perm_magn_2}) and~(\ref{eq:fields_perm_magn_3}). Subsequently, the~power switch is turned on and the current is let to flow through the coils. To~minimize the magnetic field readout disturbances induced by the sudden turn on of the current, a~wait time of 0.2~s is introduced. The~exact length of the wait depends on the characteristics of the coils of interest and needs to be determined by the user as this is highly coil-dependent ({it can 
 be facilitated by the real-time mode of the magnetometer; see Section~\ref{subsec:gui}}). After this brief waiting period, the~total magnetic field is measured and the field from coils is calculated with Equation~(\ref{eq:fields_perm_magn_1}). The~coil mode is expected to provide better results in environments where the stray magnetic field varies over time since the time between the stray field and the magnetic field measurements is shorter than in the permanent mode.

\begin{figure}[H]
\centering
\includegraphics[width=\columnwidth]{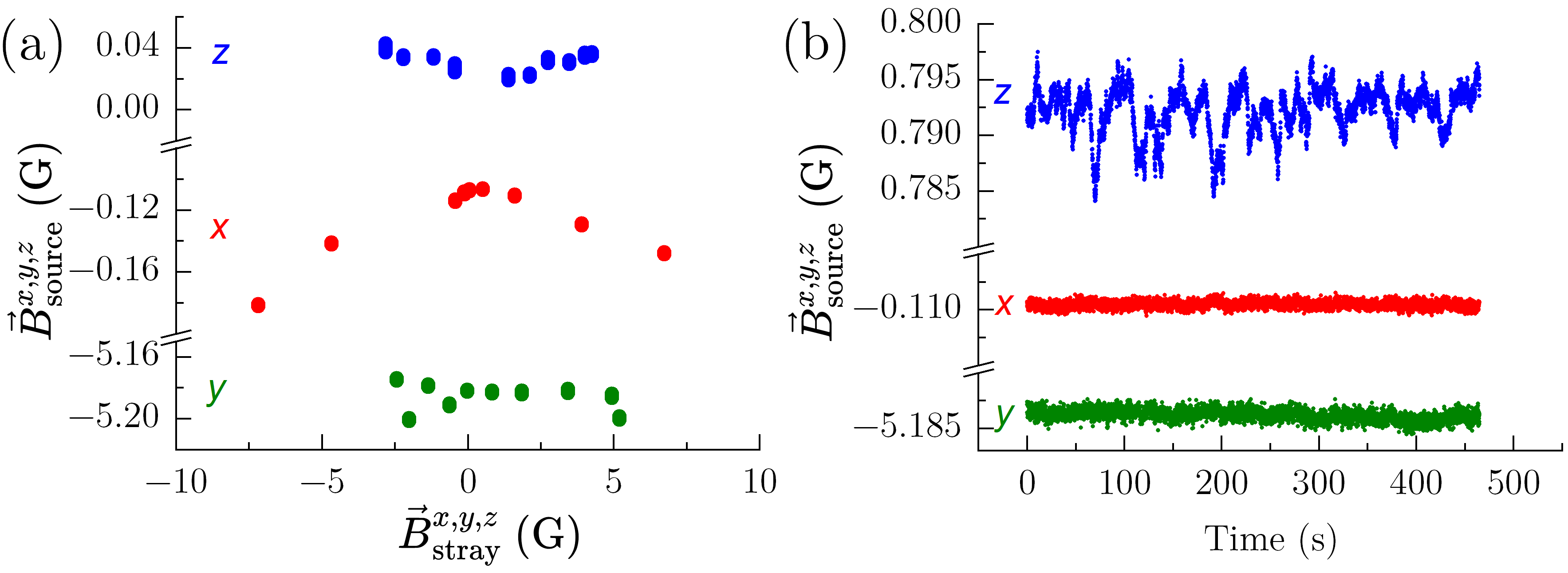}
\caption{\label{fig:source_vs_stray_vs_time} Magnetic 
 field produced by the coils measured at a fixed location for different stray fields (\textbf{a}) and as a function of time (\textbf{b}). Magnetic field projections along the sensor axes $x$, $y$ and $z$ are color-coded with red, green and blue, respectively.}
\end{figure}

In order to assess the reproducibility of the measurement of the field created by the coils, we have determined it as a function of different stray fields (Figure~\ref{fig:source_vs_stray_vs_time}a) and as a function of time for the ambient laboratory stray field (Figure~\ref{fig:source_vs_stray_vs_time}b). For~these measurements, the~sensor head has been parked in the center of one of the coils. Stray fields have been controlled by placing a small neodymium magnet in different locations in the vicinity of the coil, and, for each value, 30 data readouts of the coil field have been performed. Figure~\ref{fig:source_vs_stray_vs_time}a shows all the acquired data, resulting in apparent elongation of the symbols in the vertical direction. For~fields on the order of Earth's magnetic field, the variation in the coil field along each axis is $\sim$10~mG. The~field created by coils has not changed noticeably within 500~s interrogation time, as~can be seen in Figure~\ref{fig:source_vs_stray_vs_time}b.

We have observed that the offset of the Hall sensor changes during data acquisition, as can be seen in Figure~\ref{fig:stray_offset}b. This is most likely caused by the exposure of the sensor to the magnetic field, including spatially changing stray fields. This observation further enhances the approach to data acquisition that we have taken because~this varying offset does not influence the quantities we are interested in, i.e.,~the magnetic field from the source and stray~fields.

We have estimated the time it would take to perform a magnetic field characterization with a 1~mm resolution over a distance of 20~cm with and without averaging. For~the permanent mode, it takes 230~s~(1320~s with averaging) to complete, whereas, for the coil mode, it takes 170~s~(1260~s with averaging). For~a large number of averaged points, both acquisition modes take pretty much the same because the measurements are dominated by the time it takes to read out data from the Hall sensor. The~current turn on/turn off time is almost the same as the time it takes to move the stepper motor to the next location---which happens twice as many times in the permanent mode. 

\section{Conclusions}
We have demonstrated an automated magnetometer that can be used to characterize magnetic fields created by coils and/or permanent magnets. The~presented design is optimized for easy reproduction by inexperienced users by relying on off-the-shelf elements and 3D-printed parts. The~entire software suit needed for the operation of the device comes with a GUI that enables the control of all essential functions. If~one decides to use different hardware solutions (e.g., a different Hall sensor or a different stepper motor), the modifications of the original code should be straightforward as long as the new devices can communicate with the same protocols as are used in our~design. 

We have characterized the performance of the magnetometer and have demonstrated how the access to all three components of the magnetic field vector can be used to optimize the alignment of the coils. We have shown an important property of our design, i.e.,~the ability to extract magnetic fields produced by the source in an environment plagued by spatially and temporally varying stray fields. The~obtained field values agree with the data obtained with a commercial gaussmeter within 3$\sigma$ of its uncertainty (1\% of the \mbox{measured value).}

The $\pm8$~G saturation magnetic field of the Hall sensor we use might be too small for certain applications, in~particular for characterization of some permanent magnets that can produce fields on the order of hundreds of Gauss even several cm away from their surface. For~such high fields, an~upgrade with a easily available Hall sensor with a higher saturation threshold is not feasible. As~a solution, one could build a low-quality $\mu$-metal shield that would weaken the magnetic field but would not shield it completely. This approach should be easy to implement with thin $\mu$-metal~sheets.

It is worth noting, as~discussed in Section~\ref{sec:general_consideration}, that the polarization flipping function of the sensor is not always necessary to determine the magnetic field generated by the coils or permanent magnets, $\vec{B}_{\mathrm{coil}}$. Any method of determining the magnetic field created by the source that corrects for stray fields relies on the ability to remove the source (turn off the coils or remove permanent magnets). In~principle, the~magnetic field measurement with the coils on ($\vec{B}_{\mathrm{ON}}$) and coils off ($\vec{B}_{\mathrm{OFF}}$) would provide sufficient information to extract the value of interest, $\vec{B}_{\mathrm{coil}}=\vec{B}_{\mathrm{ON}}-\vec{B}_{\mathrm{OFF}}$. However, we aim here at the versatility of the setup such that our design can also be used to reliably measure stray fields, which can be completed only if the internal offset of the sensor can be~determined.




\vspace{6pt} {}
\authorcontributions{K.D. designed, constructed and characterized the magnetometer; M.S. initiated and supervised the project, devised the measurement plan, acquired funding, analyzed and interpreted data; M.S. and K.D. wrote the~manuscript. All authors have read and agreed to the published version of the manuscript.}

\funding{This research was funded by the Foundation for Polish Science within the Homing programme and the National Science Centre of Poland (grants No. 2016/21/D/ST2/02003, No. 2021/43/B/ST4/03326 and a postdoctoral fellowship for M.S., grant No. DEC-2015/16/S/ST2/00425).}

\dataavailability{The data presented in this study are available on request from the corresponding author (M.S.).} 


\conflictsofinterest{The authors declare no conflicts of~interest.}

\begin{adjustwidth}{-\extralength}{0cm}
\reftitle{References}



\end{adjustwidth}

%



\end{document}